\documentclass[twocolumn,notitlepage,superscriptaddress, nofootinbib,nobibnotes, aps,prd,10pt]{revtex4-1}% secnumarabic, 
\usepackage{amsmath,amssymb, fp}  
\usepackage{xcolor}  
\usepackage{soul} 
\usepackage{booktabs}
\definecolor{linkcolor}{RGB}{7,94,84}  %teal dark green
\usepackage[	colorlinks=true,    	
			pdfstartview=FitV,
			linkcolor= linkcolor,
			citecolor= linkcolor,
			urlcolor= linkcolor,
			linktoc=page,
			hyperindex=true,
			hyperfigures=true,  
			breaklinks=true]
			{hyperref}
\usepackage{dblfloatfix}    
\usepackage{balance} 
\usepackage{lastpage}  
\usepackage{mathtools}
\usepackage{epsfig,rotating,pifont}
\usepackage{graphicx}
\graphicspath{{./figures/}}	
\usepackage[caption=false]{subfig}
\usepackage{placeins}
\usepackage{ragged2e} 
\usepackage{etoolbox}
\usepackage{changepage}   % for the adjustwidth environment
\usepackage{pgfplots}
\usetikzlibrary{pgfplots.groupplots}
\pgfplotsset{compat=1.3}
\usepackage{tikz}  
\usetikzlibrary{arrows,shapes,positioning}
\usetikzlibrary{decorations.markings,decorations.pathmorphing,decorations.pathreplacing,decorations.text}
\usetikzlibrary{arrows,calc,patterns,shapes.geometric}
\usepackage{fancybox}
\usepackage{verbatim}
\usepackage{multirow}
\usepackage{titlesec}

\newsavebox\CBox

\def\rd{\textrm{d}}

\def\beq{\begin{eqnarray}}
\def\eeq{\end{eqnarray}}

\newcommand{\SL}{\mathrm{SL}}
\newcommand{\SO}{\mathrm{SO}}

\renewcommand{\sl}{{\mathfrak{sl}}}
\renewcommand{\so}{{\mathfrak{so}}}

\newcommand{\schr}{{\mathfrak{sh}}}

\def\C{\mathcal{C}}

\def\H{\mathcal{H}}

\def\O{\mathcal{O}}

\def\cR{\mathcal{R}}

\def\V{\mathcal{V}}
\def\X{\mathcal{X}}

\def\cM{\mathcal{M}}

\def\eps{\epsilon}

\def\la{\langle }
\def\ra{\rangle }

\newcommand{\R}{{\mathbb R}} % Real Numbers  % amssymb
\newcommand{\be}{\begin{equation}}
\newcommand{\ee}{\end{equation}}
\newcommand{\bea}{\begin{eqnarray}}
\newcommand{\eea}{\end{eqnarray}}
\newcommand{\bg}{\begin{gather}}

\newcommand{\bseq}{\begin{subequations}}
\newcommand{\eseq}{\end{subequations}}

\def\be{\begin{eqnarray}}
\def\ee{\end{eqnarray}}

\def\nn{\nonumber}

\def\f{\frac}

\definecolor{Green}{RGB}{147,162,153}
\definecolor{Green2}{RGB}{26,148,49}
\definecolor{BrownL}{RGB}{173,143,103}
\definecolor{Red}{RGB}{210,83,60}
\definecolor{BrownD}{RGB}{114,96,86}
\definecolor{GreyD}{RGB}{76,90,106}
\definecolor{GreyB}{RGB}{128,141,160}
\definecolor{Maroon}{RGB}{121,70,61}
\definecolor{Blue}{RGB}{148,184,210}
\definecolor{Blue2}{RGB}{108,144,170}
\definecolor{Blue3}{RGB}{42, 107, 172}
\definecolor{BB}{RGB}{128,184,220}

\newsavebox\foobox
\def\pp{\partial}
\def\tt{\tilde{t}}
\def\tx{\tilde{x}}
\def\tPsi{\widetilde{\Psi}}
\def\bPsi{\bar{\Psi}}
\def\vo{\alpha}
\def\vt{\beta}
\def\dvo{\dot{\alpha}}
\def\dvt{\dot{\beta}}
\def\po{p_{\vo}}
\def\pt{p_{\vt}}
\def\bHl{\mathbf{H}^{(\Lambda)}}
\def\bH{\mathbf{H}}
\def\w{\wedge}

\begin{document}

\title{Schr\"{o}dinger symmetry of Schwarzschild-(A)dS black hole mechanics}  
%\title{The Schwarzschild-(A)dS black hole as a many-body Schr\"{o}dinger system}  

\author{Jibril Ben Achour}
\email{j.benachour@lmu.de}
\affiliation{Arnold Sommerfeld Center for Theoretical Physics, Munich, Germany}
\affiliation{Univ de Lyon, ENS de Lyon, Laboratoire de Physique, CNRS UMR 5672, Lyon 69007, France}
\author{Etera R. Livine}
\email{etera.livine@ens-lyon.fr}
\affiliation{Univ de Lyon, ENS de Lyon, Laboratoire de Physique, CNRS UMR 5672, Lyon 69007, France}
\author{Daniele Oriti}
\affiliation{Arnold Sommerfeld Center for Theoretical Physics, Munich, Germany}

\date{February 13, 2023} 
 
\begin{abstract}
%\vspace{-1pt}
%\vspace*{-8pt}
%$\,$

We show that the dynamics of Schwarzschild-(A)dS black holes admits a symmetry under the 2d Schr\"{o}dinger group, whatever the sign or value of the cosmological constant.
This is achieved by reformulating the spherically-symmetric reduction of general relativity as a 2d mechanical system with a non-trivial potential controlled by the cosmological constant, and explicitly identifying the conserved charges for black hole mechanics.
We expect the Schr\"{o}dinger symmetry to drive the dynamics of quantum Schwarzschild-(A)dS black holes.
This suggests that Schr\"{o}dinger-preserving non-linear deformations (of the Gross-Piteavskii type) should capture universal quantum gravity corrections to  the black hole geometry. Such scenario could be realized in condensed matter analogue models.

%The Schr\"{o}dinger group is the key symmetry of classical non-relativistic mechanics which is preserved by quantization. We thus expect it to drive the dynamics of quantum Schwarzschild-(A)dS black holes.
%%
%Assuming that this symmetry must indeed be preserved at the quantum level, this suggests that Schr\"{o}dinger-preserving non-linear deformations (of the Gross-Piteavskii type) should capture universal quantum gravity corrections to  the black hole geometry. Such scenario could be realized in condensed matter analogue models.

\end{abstract}

\maketitle  

\makeatletter

\makeatother

%%%%%%%%%%
\section*{Introduction}
%%%%%%%%%%

Black holes are iconic predictions of General Relativity which stand as a fantastic window to unravel the fundamental structure of spacetime. Indeed, the laws of black hole mechanics and their thermodynamical interpretation have revealed that they are equipped with an entropy and a temperature \cite{Bekenstein:1973ur, Bardeen:1973gs}. It follows that black holes can be understood as many-body systems built from the collective behavior of (still unknown) microscopic degrees of freedom. Such thermodynamical point of view on gravitational systems has been widely extended since then, to cosmological spacetime, causal diamonds and light cones geometries. The key challenges in completing this picture are on the one hand, to identify the nature of these microscopic degrees of freedom, and on the other hand, to understand the emergence of classical geometries from such microscopic description.
While there might be different ways to encode the microscopic degrees of freedom depending on the chosen model or theory, one expects that their dynamics, and thus the emergence of spacetime in the continuum hydrodynamical approximation, to be governed by universal symmetries.

Dualities between gravitational and condensed matter systems, for which the mean-field approximation methods are well under control, provides a powerful avenue to shed light on these issues. Such mapping naturally emerged in the non-relativistic regime of holographic gauge/gravity dualities such as the AdS/CFT correspondence.
%Of course, the key motivation behind such non-relativistic regime stands as the possibility to experimentally use the dictionary in analogue condensed matter systems. 
In view of the prominent role played by the Schr\"{o}dinger equation and its non-linear extensions in non-relativistic physics, an important effort has been devoted to construct cold atoms/gravity correspondence based on the Schr\"{o}dinger group \cite{Son:2008ye, Balasubramanian:2008dm}. Concretely, non-relativistic holography relates manifolds with Schrodinger isometries to non-relativistic CFT living on their boundary  \cite{Taylor:2008tg, Goldberger:2008vg, Rangamani:2008gi}. Condensed matter systems enjoying such non-relativistic conformal symmetry are characterized by an anisotropic scaling invariance  of the spacetime coordinates of the form
\be
t \rightarrow \lambda t \;, \qquad x^i \rightarrow \lambda^z x^i
\ee
where $z$=2 is the critical exponent. Such invariance appears in a variety of contexts, from strongly correlated fermions, vortices, monopoles, compressible fluid mechanics and in  Bose-Einstein condensates. In particular, this conformal  symmetry is realized for suitable non-linear Schr\"{o}dinger equations describing ultra cold atoms gases, such as the Gross-Piteavskii condensate and the Tonks-Girardeau gas \cite{Kolo,Ghosh:2001an}. While the construction of dualities between such condensed matter systems and gravity has mostly been investigated in the framework of non-relativistic holography, it seems that dictionaries between non-linear Schr\"odinger and gravity could be identified based directly on the shared symmetries of the two classes of systems.

\smallskip

The goal of this short paper is to develop this storyline for Schwarzschild-(A)dS black holes. Concretely, we consider the spherically-symmetric stationary reduction of general relativity, that can be called more descriptively Schwarzschild-(A)dS black hole mechanics. And we show that this reduced gravitational model admits a symmetry
%has an in-built invariance
under the 2d Schr\"{o}dinger group, whatever the sign and value of the cosmological constant.
This is achieved by explicitly identifying the conserved charges generating this symmetry.
We expect this symmetry to be conserved when quantizing the system. Indeed, standard quantization is meant to identify suitable representations (in the mathematical sense) of the symmetry group. Then breaking a classical symmetry at the quantum level usually reveals a deep physical phenomenon, with strong experimental signatures such as anomalies, phase transitions, emergent collective  modes or new propagating degrees of freedom.
Here the 2d Schr\"{o}dinger group is the symmetry group of classical 2d mechanics, consisting of the Galilean relativity transformations plus conformal transformations. It is conserved by standard quantum mechanics.
One can thus expect this symmetry to also be preserved when considering quantum gravity corrections to the black hole geometry.
Deforming it, or breaking it, would signal a departure from the standard quantization scheme (e.g. such as non-commutative deformations \cite{Banerjee:2005zt}) and/or new physics for quantum black holes.

This sets a strong criterion to discriminate between regularized black hole metric proposals in quantum gravity phenomenology.
For instance, assuming that quantum Schwarzschild-(A)dS black holes could be generally modeled as a non-linear extension of 2d quantum mechanics with a self-interaction between black hole quanta, preserving the Schr\"{o}dinger group symmetry fixes the self-interaction term to be in $\psi^{4}$, thus implying a universal UV behavior to quantum black holes and the existence of a dictionary between black hole quantum mechanics and the Gross-Pitaevskii equation.
This scenario is especially interesting with respect to the possibility of imagining a new type of analogue quantum black hole systems, e.g. with Bose-Einstein condensates,
%or more generally with condensed matter systems,
based on an exact mapping between dynamical conserved charges and not anymore on mimicking the Schwarzschild spacetime metric as for sonic black holes.
This possibility could then be extended to a large class of cosmological dynamics following the symmetry and conserved charge analysis of %\cite{BenAchour:2022fif, Geiller:2022baq}.
%\cite{BenAchour:2019ufa, BenAchour:2020xif, Achour:2021lqq, BenAchour:2022fif, Geiller:2022baq}.
\cite{BenAchour:2022fif, Geiller:2022baq,Achour:2021lqq}.

We start by reviewing the Schr\"odinger symmetry of classical mechanics, which encodes its invariance under Galilean and conformal transformations, and showing that it is indeed preserved under quantization. The Casimirs of the Schr\"odinger group, initially vanishing at the classical level, acquires non-zero values in quantum mechanics and reflect the extra degrees of freedom represented by the wave-function dressing the classical system.
Although this material is not new, these aspects are often not emphasized. They are nevertheless crucial to the identification of the symmetry of the black hole dynamics.
Then moving to Schwarzschild-(A)dS black holes, the spherically-symmetric reduction of general relativity can be written as a mechanical system, with a non-trivial potential given by the cosmological constant term, and we show that this potential does not spoil the invariance under the Schr\"odinger symmetry. We underline an important difference with standard mechanics: the non-positive signature of the kinetics, which actually comes from studying a relativistic system. In particular, we identify the black hole mass (thus its energy) as a boost generator within the symmetry algebra, thereby echoeing discussions about the modular Hamiltonian for black hole thermodynamics. Finally, we argue that this should be a key symmetry for quantum black holes and we discuss its relevance for quantum gravity.

\section{Schr\"{o}dinger symmetry and Galilean relativity}
%%%

Let us start with reviewing the algebra of conserved charges for the classical mechanics of a free particle in $d$ spatial dimensions, driven by the action:
\be
S[t,x^{a}]=\f m2\int\rd t \, \dot{x}^{a}\dot{x}_{a}\,,
\ee
where $m$ is the particle's mass and the index $a$ runs from 1 to $d$. The canonical analysis defines the conjugate momentum and Poisson bracket,
\be
p_{a}=m\dot{x}_{a}\,,\quad
\{x^{a},p_{b}\}=\delta^{a}_{b}\,,
\ee
and the Legendre transform gives the Hamiltonian,
\be
S[t,x^{a}]=\int\rd t\,\left[
p_{a}\dot{x}^{a}-H
\right]
\quad\textrm{with}\,\,
H=\f1{2m}p_{a}p^{a}\,.
\ee
By Noether theorem, symmetries are generated by conserved charges. In general, those conserved charges can depend explicitly on time and satisfy 
\be
\rd_{t}\O=\pp_{t}\O+\{\O,H\}=0
\,.
\ee
The algebra of conserved charges for the free particle is well known. It leads to the Schr\"odinger algebra, which reflects the free particle's invariance under the Galilean transformations and conformal transformations. This construction is crucial, because this is the maximal symmetry preserved by the quantization.
In more details, a first set of conserved charges consists in the momentum $p_{a}$, the Galilean boost generator $b_{a}$ and the angular momentum $j_{ab}$,
\be
b_{a}=\f1m\big{[}mx_{a}-tp_{a}\big{]}\,,\quad
j_{ab}=x_{a}p_{b}-x_{b}p_{a}
\,,
\ee
which satisfy the Galilean algebra
\begin{align}
&\{p_{a},p_{b}\}=\{b_{a},b_{b}\}=0\,,\quad
\{b_{a},p_{b}\}=\delta_{ab}\,,
%\{b_{a},p_{b}\}=m\delta_{ab}\,,
\\
&\{j_{ab},p_{c}\}=\delta_{ac}p_{b}-\delta_{ab}p_{c}\,,
\quad
\{j_{ab},b_{c}\}=\delta_{ac}b_{b}-\delta_{ab}b_{c}\,,
\nn\\
&\{j_{ab},j_{cd}\}=\delta_{ac}j_{bd}-\delta_{ad}j_{bc}-\delta_{ad}j_{bc}+\delta_{bd}j_{ac}
\,.
\nn
\end{align}
The momentum $p_{a}$ generates the symmetry under space translations $x^{a}\mapsto x^{a}+w^{a}$, while the angular momentum $j_{ab}$ generates the symmetry under $\SO(d)$ space rotations. The vector $b_{a}$ depends explicitly on the time $t$, it is an evolving constant of motion, indicating the initial condition (at $t=0$) for the particle position. It can be interpreted as an extra component of the angular momentum with respect to a pair of conjugate variables $(x^{0},p_{0})=(t,m)$. It generates the symmetry under translation by a fixed speed,
\be
x^{a}\mapsto x^{a}+v^{a}t\,,\qquad
p^{a}\mapsto p^{a}+mv^{a}\,.
\ee
Together, $(p_{a},b_{a},j_{ab})$ encode the Galilean relativity of the free classical particle. To these, we add three other conserved charges $q_{\mu}$, defined as
\begin{align}
q_{+}&=mH\,,\\
2q_{0}&=D-2Ht\,,\nn\\
2mq_{-}&=mx^ax_{a}-2tD+2t^{2}H\,,\nn
\end{align}
where we have introduced the dilatation generator $D=x^{a}p_{a}$. These three observables form a $\sl(2,\R)$ Lie algebra,
\be
\{q_0, q_{\pm}\}  = \pm  q_{\pm}
\,,  \quad
\{q_{+}, q_{-} \} = -2 q_0
\,,
\ee
and generate the conformal symmetry of the free particle: $q_{+}\propto H$ generates time translations, $q_{0}$ is the initial condition for $D$ and generates inverse rescalings of the position and momentum, finally $q_{-}$ gives the initial condition for the squared distance $x^{2}$ and generates special conformal transformations.
This conformal symmetry is a universal feature of mechanical systems, leading for instance to the conformal structure of the Hydrogen atom spectrum (e.g. \cite{Bars:1998pc}).

The $\sl(2,\R)$ does not commute with the Galilean sector; the non-vanishing brackets are:
\be
\begin{split}
\{q_{0},p_{a}\}&=+\tfrac{1}2{p_{a}}\,,\\
\{q_{0},b_{a}\}&=-\tfrac{1}2{b_{a}}\,,
\end{split}
\quad
\begin{split}
\{q_{-},p_{a}\}&=+b_{a}\,,\\
\{q_{+},b_{a}\}&=-p_{a}\,,
\end{split}
\ee
%\be
%\begin{split}
%\{Q_{0},p_{a}\}&=+\tfrac{1}2{p_{a}}\,,\\
%\{Q_{0},b_{a}\}&=-\tfrac{1}2{b_{a}}\,,
%\end{split}
%\quad
%\begin{split}
%\{Q_{+},p_{a}\}&=0\,,\\
%\{Q_{+},b_{a}\}&=-p_{a}\,,
%\end{split}
%\quad
%\begin{split}
%\{Q_{-},p_{a}\}&=b_{a}\,,\\
%\{Q_{-},b_{a}\}&=0\,.
%\end{split}
%\ee
Putting all the conserved charges together, this algebra is known as the d-dimensional Schr\"{o}dinger algebra $\schr(d)$,
\be
\schr(d) = (\sl(2,\mathbb{R}) \oplus \so(d)) \oplus_{s} (\mathbb{R}^d \oplus \mathbb{R}^d)\,,
%\schr(d) = (\sl(2,\mathbb{R}) \times \so(d)) \ltimes (\mathbb{R}^d \times \mathbb{R}^d)
\ee
where $\oplus_{s}$ denotes a semi-direct sum, where the $\sl(2,\mathbb{R})$ sector generated by the $q$'s and the $\so(d)$ sector generated by the $j$'s act non-trivially on the $\mathbb{R}^d \oplus \mathbb{R}^d$ sector consisting in the $p$'s and $b$'s.
Once exponentiated, these charges give the Schr\"odinger symmetry group,
\be
\textrm{Sh}(d)=(\SL(2,\mathbb{R}) \times \SO(d)) \ltimes (\mathbb{R}^d \times \mathbb{R}^d)\,.
\ee
This is the key symmetry group of mechanics preserved by quantization.

\smallskip

An important remark is that, while there are $2d$ independent variables in the phase space, given by the pairs $(x^{a},p_{a})$, we have identified $3+d(d-1)/2+2d$ conserved charges. This means that these constants of motion are clearly redundant and that there exists relations between them. These relations are nevertheless not linear, and it is important to keep in mind that a non-linear combination of Lie algebra generators lays by definition out of that Lie algebra: the symmetry transformations generated by a conserved charge or a power of that charge are a priori not the same.

%Charges non-independent and clearly redundant. but diff between Lie algebra and algebra. Relation between charges. Schr\"odinger Casimirs.
Let us focus here on the two-dimensional case $d=2$. A more systematic treatment for arbitrary dimension can be found in \cite{Alshammari:2017jky}.
For $d=2$, the angular momentum has a single component $j\equiv j_{12}$.
A first relation expresses it in terms of the two pairs of constants of motion $(b^{a},p_{a})$,
\be
\C_{2}\equiv b\w p-j=0
\,\,\;\textrm{with}\,\,\,b\w p=\big{(}b_{1}p_{2}-b_{2}p_{1}\big{)}
\,,
\ee
which reflects that the $b_{a}$'s are simply the evolving constants of motion for the positions $x^{a}$. This actually is the quadratic Casimir of the Schr\"odinger algebra: it commutes with all the Schr\"odinger charges, and thus is invariant under translations, boosts, rotations and conformal transformations.
Another set of conditions resulting from the expressions of the charges in terms of $x$'s and $p$'s gives the conformal charges in terms of the boost charges and momenta:
\be
q_{+}=\f{p^{2}}2
\,,\quad
q_{0}=\f{b^{a}p_{a}}2
\,,\quad
q_{-}=\f{b^{2}}2
\,.
\ee
But these relations are not invariant under conformal transformations. Another important relation is the balance equation giving the $\sl(2,\R)$ Casimir in terms of the angular momentum:
\be
q_{+}q_{-}-q_{0}^{2}=\tfrac14 j^{2}
\,,
\ee
but it is not invariant under translations or boosts. It is nevertheless possible to repackage these relations in terms of the cubic Casimir of the Schr\"odinger algebra,
\begin{eqnarray}
\C_{3}\equiv&\,\,&
q_{0}^{2}-q_{+}q_{-}+\tfrac14 j^{2}\\
&&+\tfrac{b^{2}}2q_{+}+\tfrac{p^{2}}2q_{-}-b^{a}p_{a}q_{0}-\tfrac{b\w p}2j
=0\,,\nn
\end{eqnarray}
which is appropriately invariant under all Schr\"odinger symmetries.

%{\bf Relations and Casimirs changed by quantization. Reveal dressing of classical particle with quantum fluctuations}
%
Although the Schr\"odinger symmetry algebra is preserved by the quantization, and even characterizes the quantization procedure, these relations and vanishing Casimir conditions, $\C_{2}=\C_{3}=0$, are not valid at the quantum level anymore. Their non-zero values actually encode the dressing of the classical particle with quantum fluctuations and reveal the infinite tower of new degrees of freedom when upgrading the classical variables $(x^{a},p_{a})$ to a wave-function $\Psi(x^{a})$.

The goal of the present letter is to show that the dynamics of (spherically symmetric) black holes in general relativity is also driven by the same Schr\"odinger symmetry charges, as pointed out in \cite{BenAchour:2022fif}, to extend those previous results to include a non-vanishing cosmological constant, and to discuss its role in describing quantum black holes.

%%%
\section{Symmetry of Quantum Mechanics}
\label{sec:QM}
%%%

Before moving on to black holes, we discuss the fate of the Schr\"odinger symmetry in standard non-relativistic quantum mechanics.
%
%Let us review the conserved charges, and the corresponding symmetry transformations, for Schr\"{o}dinger equation in standard non-relativistic quantum mechanics.
%
We consider the free Schr\"{o}dinger system in $d$-spatial dimension defined by the field theory Lagrangian:
\begin{align}
S [\Psi, \bar{\Psi}] =\int \rd t \rd^d x
\,\left[
i \hbar\bar{\Psi} \partial_t \Psi - \frac{\hbar^{2}}{2m}  \partial_a \Psi \partial^{a} \bar{\Psi}
\right]
\,.
%\,,\nn 
\end{align}
%where Planck constant  has been set to $\hbar=1$.
%
The resulting field equation is the Schr\"{o}dinger equation:
\be
i \partial_t \Psi =-\f\hbar{2m} \partial_a \partial^a \Psi
\,,
\ee
which gives the equation of motion for the wave-function $\Psi$ in the $x$-polarization.
%
%Using a polarization where the wave function depend on the spatial coordinates, i.e $\Psi = \Psi(\vec{x})$,
%
The canonical analysis of this action gives the pair of conjugate variables,
\be
\{ \Psi(x), \bar{\Psi}(y)\} = \tfrac1{i\hbar}\,\delta^{(d)}(x-y)\,,
\ee
and the field theory Hamiltonian,
% which is also the expectation value of the Hamiltonian operator,
\be
H  = -\f{\hbar^{2}}{2m}\int \rd^d x \;\bar{\Psi} \partial_a \partial^a \Psi 
%Q_{+}= \int \rd^d x \left( \bar{\Psi} \partial_i \partial^i \Psi \right)
\,.
\ee
We  introduce the  probability integral $n=\int \rd^d x \,\bar{\Psi} \Psi$, also understood as the number of particles, and the average position and momentum,
\be
%n  = \int \rd^d x \,\bar{\Psi} \Psi
%\,,\quad
X^a = \int \rd^dx\,  \bar{\Psi} x^a\Psi 
\,,\quad
P_a  = {-i\hbar}\int \rd^d x  \, \bar{\Psi}  \partial_a\Psi\,, 
\ee
as well as the quadratic moments of the wave function,
\beq
J_{ab} & = &-i\hbar\int \rd^d x \; \bar{\Psi} \left(  x_a  \partial_b -  x_b \partial_a  \right) \Psi 
\,,\\
D & =&\frac{-i\hbar}{2} \int \rd x^d \; \bar{\Psi} \left(  x^a  \partial_a +  \partial_a  x^a \right) \Psi
\,,\\
\X&=&\int \rd^dx\,  \bar{\Psi} x^ax_{a}\Psi 
\,.
\eeq
The angular momentum $J_{ab}$, the expectation value $C$  of the  dilatation generator $\vec{x}\cdot\vec{p}$ and the position uncertainty $\X$  characterize the shape of the wave packet.

The integrals, $n$, $P_{a}$ and $J_{ab}$, have vanishing Poisson brackets with the Hamiltonian,
and are thus constants of motion, $\{n,H\}=\{P_{a},H\}=\{J_{ab},H\}=0$.
%\be
%\{n,H\}=\{P_{a},H\}=\{J_{ab},H\}=0
%\,.
%\ee
%
As for classical mechanics, we introduce the evolving position observable:
\be
B_{a}  & =  X_{a} - \f tmP_{a}
\,,\quad
\rd_{t}B_{a}=\pp_{t}B_{a}+\{B_{a},H\}=0\,.\,\,
\ee
We compute the Poisson brackets between those observables,
\begin{align}
\label{n}
& \{ B_a , P_b \} = \delta_{ab} n  \,,\\
& \{J_{ab}, P_c\} =  \delta_{ac} P_b - \delta_{bc} P_a \,,\nn\\
&  \{J_{ab}, B_c\} = \delta_{ac} B_b - \delta_{bc} B_a \,,\nn\\
&  \{J_{ab}, J_{cd}\} = \delta_{ac} J_{bd} -\delta_{bc} J_{ad} -\delta_{ad} J_{bc} +\delta_{bd} J_{ac} \nn\,,
\end{align}
which form a centrally extended Galilean algebra, with the number of particles $n$ as the central charge.
We complete this set of conserved charges with the constants of motions encoding the evolution of the quadratic quantum uncertainty:
\begin{align}
Q_{+}&=mH
\,, \\
2Q_0  & = D - 2 H t \nn
\,, \\
2mQ_{-} &=m \X - {2} tD+2 t^{2}H
\,.\nn
\end{align}
The evolving constants of motion $Q_{0}$ and $Q_{-}$ are the initial conditions at $t=0$, respectively for  the observable $D$ and the position spread $\X$. Their explicit time dependence exactly compensates their non-vanishing brackets with the Hamiltonian.
As expected, these form a $\sl(2,\mathbb{R})$ algebra, 
\be
\{Q_0, Q_{\pm}\}  = \pm  Q_{\pm}
\,,  \quad
\{Q_{+}, Q_{-} \} = -2 Q_0
\,,
\ee
whose Casimir is $\C_{\sl}=Q_{0}^{2}-Q_{+}Q_{-}$.
This is the quadratic uncertainty algebra of \cite{Livine:2022vaj}.
The remaining non-vanishing bracket are given by
\begin{align}
\{Q_{0},P_{a}\}=+\tfrac12P_{a} \,,\quad \{Q_{-},P_{a}\}=+B_{a}\,,\\
\{Q_{0},B_{a}\}=-\tfrac12B_{a} \,,\quad \{Q_{+},B_{a}\}=-P_{a}\,.\nn
\end{align}
We recognize the same d-dimensional Schr\"{o}dinger algebra $\schr(d)$ as for classical mechanics,
\be
\schr(d) = (\sl(2,\mathbb{R}) \oplus \so(d)) \oplus_{s} (\mathbb{R}^d \oplus \mathbb{R}^d)
\,.
%\schr(d) = (\sl(2,\mathbb{R}) \times \so(d)) \ltimes (\mathbb{R}^d \times \mathbb{R}^d)
\ee
%where $\oplus_{s}$ denotes a semi-direct sum, where the $\sl(2,\mathbb{R})$ sector generated by the $Q$'s and the $\so(d)$ sector generated by the $J$'s act non-trivially on the $\mathbb{R}^d \oplus \mathbb{R}^d$ sector consisting in the $P$'s and $B$'s.
%
%Let us make several important remarks to conclude this review of the Schr\"odinger symmetry group.
%
%First, the Schr\"odinger charges, $(P_{i},B_{i}, J_{ij},Q_{0},Q_{\pm})$, can also be written on the phase space of a classical free particle, as linear and quadratic combinations of the canonical pair $(x^{i},p_{i})$. In that case, the linear observables $(P_{i},B_{i})$ fully characterize the classical trajectory and position of the particle, and the quadratic observables $(J_{ij},Q_{0},Q_{\pm})$ do not carry more information about the system. However, in the quantum theory, the wave-function $\Psi(x)$ contains infinitely more information than the classical position and momentum. Indeed, the charges $(J_{ij},Q_{0},Q_{\pm})$ are now independent from linear observables $(P_{i},B_{i})$ and reflect the (ellipsoidal) shape of the wave-packet. They are legitimate degrees of freedom, representing the quantum fluctuations on top of the (semi-)classical motion. From this perspective, the Schr\"{o}dinger charges provide an algebraic relation between classical and quantum degrees of freedoms, as illustrated by the analysis in \cite{Livine:2022vaj}.
%
The important difference with classical mechanics is that the Schr\"odinger Casimirs do not vanish anymore. This reveals a tower of extra degrees of freedom. Indeed, the Schr\"odinger charges for the classical particle could all be written as polynomials in the canonical position and momentum. This is no longer the case in quantum mechanics. The wave-function $\Psi$ contains infinitely more information than the classical position and momentum: the charges $(J_{ij},Q_{0},Q_{\pm})$ are now independent from the linear observables $(P_{i},B_{i})$ and encode the shape of the wave-packet, they are legitimate degrees of freedom, representing the quantum fluctuations on top of the classical motion.

To be more precise, we can look into the $d=2$ case. A non-zero quadratic Casimir reveals an extra contribution to the angular momentum,
\be
\C_{2}= \la \hat{x}_{1}\ra\la\hat{p}_{2}\ra-\la\hat{x}_{2}\ra\la\hat{p}_{1}\ra-n\la J_{12}\ra\,\ne0\,,
\ee
which actually means that the quantum state $\Psi$ carries non-trivial correlation and entanglement between the two directions $x_{1}$ and $x_{2}$. Similarly, the cubic Casimir $\C_{3}$ relates the $\sl_{2}$ Casimir for the conformal symmetry to the Galilean generators. The fact that it does not vanish anymore, and that it can take arbitrary values, reflects that the (quadratic) quantum uncertainty - the spread of the wave packet - measured by the $Q$'s can evolve independently from the classical degrees of freedom $X^{a},P_{a}$. From this perspective, non-zero values of the Schr\"odinger Casimirs, $\C_{2}\ne0$, $\C_{3}\ne0$, are witnesses of the quantumness of the system.

\medskip

Once exponentiated, these conserved charges generate symmetries of the system according to Noether's theorem.
%
%According to Noether's theorem, these conserved charges generate symmetries of the system. The infinitesimal symmetry transformation associated to a conserved charge $\O$ is given by the Poisson bracket $\delta_{\O}\Psi= \{\O,\Psi\}$
%\be
%\delta_{\O}\Psi= \{\O,\Psi\}\,,
%\ee
%and finite symmetry transformations are obtained by exponentiating this infinitesimal variation.
%
%The number of particles $n$ generates phase transformation of the wave-function, the momentum $P_{i}$ generates spatial translations, the vector $B_{i}$ generate Galilean boosts, the Hamiltonian $Q_{+}$ generates time translation  the charge $Q_{0}$ generates time dilatations.
%
This gives the Schr\"odinger group,
\be
\textrm{Sh}(d)=(\SL(2,\mathbb{R}) \times \SO(d)) \ltimes (\mathbb{R}^d \times \mathbb{R}^d)\,,
\ee
 identified as the maximal symmetry group of the free Schr\"{o}dinger equation by Niederer in \cite{Niederer}. We catalogue, in the table \ref{table:sym}, the various symmetry transformations.
\begin{table}[!]
\caption{\label{table:sym}
Schr\"odinger conserved charges}
\vspace*{2mm}
\begin{tabular}{|l|l|}
  \hline
charge & symmetry  \\
  \hline
 $n$  & phase transformation\\
$P_{a}$ & space translations \\
$B_{a}$ & Galilean boosts \\
$Q_{+}\propto H$\,\, & time translation \\
$Q_{0}$ & time dilatation \\
$Q_{-}$ & special conformal \\
  \hline
\end{tabular}
\end{table}
While phase multiplication, translations and boosts are usual transformations, it is instructive to give a closer look at the conformal transformations.
Indeed, these are not mere rescalings. They are non-trivial symmetry transformations, creating a complex phase factor, affecting the complex width of Gaussian wave-packets, thus leading to physical effects.
More precisely, these are given by time reparameterization, with a non-trivial rescaling of the space coordinates, following e.g. \cite{Lidsey:2018byv},
\be
t\mapsto \tt=f(t)
\,,\quad
x_{a}\mapsto \tx_{a}=\dot{f}(t)^{\f12}x_{a}\,,
\ee
and both a conformal rescaling and a non-trivial phase for the wave-function,
\be
\label{eq:conformal}
\Psi\mapsto
\tPsi(\tt,\tx_{a})
=
\dot{f}(t)^{-\f d4}\,
e^{i\f m4\f{\ddot{f}}{\dot{f}}x_{a}x^{a}}\,
\Psi(t,x_{a})
\,,
\ee
which leads to the following transformation of the action,
\be
S[\tt,\tx,\tPsi]
=
S[t,x,\Psi]
-\f m4\int \textrm{Sch}[f](t)x_{a}x^{a}\Psi\bar{\Psi}\,,
\ee
with the Schwarzian derivative of the reparametrization function,
\be
\textrm{Sch}[f]
=
%\pp_{t}h
\dot{h}
-\f12h^{2}
%\pp_{t}\left(\f{\ddot{f}}{\dot{f}}\right)
%-\f12\left(\f{\ddot{f}}{\dot{f}}\right)^{2}
\,,\quad\textrm{with}\,\,
h=\ddot{f}/\dot{f}
\,.
\ee
This is a symmetry as soon as the Schwarzian derivative vanishes, i.e. when $f$ is a Mo\"ebius transformations,
\be
\textrm{Sch}[f]
=
0
\Leftrightarrow
f(t)=\f{\alpha t+\beta}{\gamma t+\delta}
\,.
\ee
This is the  $\SL(2,\R)$ symmetry group generated by the three conserved charges $Q_{0}$ and $Q_{\pm}$, as can be directly checked by looking at infinitesimal Mo\"ebius transformations.

\medskip

The purpose of the present work is to show that this Schr\"{o}dinger symmetry also controls black hole dynamics in general relativity.
This underlines the universality of the Schr\"{o}dinger charges, but also provides a direct bridge between black hole mechanics and quantum mechanics, which should shed clarifying light on the quantization of black holes. 

%We now turn to demonstrate the existence of the Schr\"{o}dinger symmetry in black hole mechanics.

%%%
%\section{Black Hole Mechanics}
\section{Schwarzschild-(A)dS Black Hole Mechanics}
%%%

We now turn to the main proof-of-concept model for general relativity,
%
%In the following, we shall focus on the central toy model for black hole physics,
namely the eternal Schwarzschild-(A)dS black hole.
%
%Let us start by reviewing the action driving the mechanics and dynamics of the 
%As a first step, we present the mechanical system encoding the
%geometry of the black hole, as presented in \cite{Achour:2021dtj}.
%
The action driving the dynamics of the geometry of the black hole is obtained by symmetry reduction and gauge-fixing from the vacuum Einstein-Hilbert-$\Lambda$ action
\be
\label{full}
S[g] =\f1{\ell^2_P} \int_{\cM} \rd^4 x \,\sqrt{|\det g|}\,\Big{[}\cR - 2\Lambda \Big{]}
\,,
%S[g] = \int_{\cM} \rd^4 x \frac{\cR - 2\Lambda}{\ell^2_P} + \int_{\partial \cM} \rd^3y \sqrt{h} K
\ee
where $\ell_P$ is the Planck length.
Boundary terms do not play any relevant role in the present analysis.
We consider a static spherically symmetric manifold $\cM = \mathbb{R}\times \Sigma_{\epsilon}$ with line element
\be
\rd s^2 = \epsilon \left(- N^2(r) \rd r^2 +  \gamma_{tt} (r) \rd t^2 \right) +  \gamma_{\theta\theta}(r) \rd \Omega^2
\,,
\ee
where $\gamma_{ij}(r)$ is the induced metric on the constant $r$ hyerpsurfaces $\Sigma_{\epsilon}$, and $\rd \Omega^2 = \rd \theta^2 + \sin^2{\theta} \rd \varphi^2$ is the standard 2-metric on the angular sector.
The parameter $\epsilon =\pm 1$ allows to deal with both interior and exterior of the black hole using the same formalism. Our conventions are naturally adapted to the case $\eps=+$ corresponding to the black hole interior: the coordinate $r$ is time-like, and the radial metric component $N(r)$ plays the role of the lapse between hypersurface. The case $\eps=-$ corresponds to the exterior region of the black hole where $r$ is a space-like coordinate and $t$ is time-like.
%
%Here $N(r)$ is the lapse, $\gamma_{ij}(r)$ is the induced metric on $\Sigma_{\epsilon}$ and $\rd \Omega^2 = \rd \theta^2 + \sin^2{\theta} \rd \varphi^2$. The parameter $\epsilon =\pm 1$ allows one to switch the nature of the foliation from timelike, i.e. $\epsilon =-1$, to spacelike, i.e. $\epsilon = +1$. Changing $\epsilon$ further induces a change of the constant coordinate labelling the foliation, i.e. $(\Sigma_{+}, r)$ and $(\Sigma_{-}, t)$. This will allow us to treat the interior and exterior of the black hole at once.

We decompose the metric components as 
\be
\gamma_{tt} := {2\vt(r)}/{\vo(r)} \;, \qquad \gamma_{\theta\theta} := \ell^2_s \vo(r)\;,
%\gamma_{tt} := \frac{2\vt(r)}{\vo(r)} \;, \qquad \gamma_{\theta\theta} := \ell^2_s \vo(r)\;,
\ee
where we introduce a fiducial length scale $\ell_s$ defining the dimensionful unit for the  $2$-sphere radius. Evaluating the full  Einstein-Hilbert-$\Lambda$ action on this metric ansatz gives the reduced action encoding the dynamics of the black hole geometry \cite{BenAchour:2022fif,Achour:2021dtj}:
\be
S_{\epsilon}[\vo,\vt]= \epsilon c \ell_P \int \rd \tau \left[ \frac{\epsilon }{\ell^2_s} - \frac{\epsilon \vo}{\ell^2_{\Lambda}}+  \frac{\vt \dvo^2- 2 \vo \dvo \dvt}{2\vo^2}  \right]
,\;\;
%\;\;\; \;\; \;\;
\ee
where we have introduced a field-rescaled radial coordinate $\tau$ defined by:
\be
\rd \tau = \sqrt{\f{2\vt}{\vo}}N(r)\rd r\,,
\ee
and the dot denotes the derivative with respect to $\tau$.
The length scale $\ell_{\Lambda} = 1/\sqrt{\Lambda}$ encodes the cosmological constant. The dimensionless constant $c$ comes from restricting the range of spatial integration to a bounded region of the hypersurface  $\Sigma_{\epsilon}$. Indeed, the metric being homogeneous, the integration over the non-compact 3-manifold automatically yields an infinite result. This is naturally resolved by introducing an infra-red cut-off $\ell_{0}$ for the coordinate $t$. This gives:
\be
\label{c}
c = \frac{1}{\ell^3_p} \int^{t_f}_{t_i} \rd t \oint \ell^2_s \rd \Omega = 4\pi\frac{\ell_0 \ell^2_s}{\ell^3_p}\,,
\ee
as the ratio between the IR scale and the UV scale of the system. 

%The profiles of the fields $(\vo, \vt)$ are then dictated by the symmetry reduced Einstein-Hilbert-$\Lambda$ action.
%At this level, the reduced action is still gauge-invariant w.r.t to the reparametrization of the $r$-coordinate. Fixing this gauge further reduces the action to the one of a mechanical system.
%This provides the reduced action for black hole mechanics we shall investigate in this work. 

%Choosing the lapse such that $N(r) = V_1(r)/(2V_2(r))$ and denoting the corresponding radial coordinate $\rd \tau = N(r)\rd r$, the full action (\ref{full}) reduces to
%\be
%S_{\epsilon}[V_1, V_2]= \epsilon c \ell_P \int \rd \tau \left[ \frac{\epsilon }{\ell^2_s} - \frac{\epsilon V_1}{\ell^2_{\Lambda}}+  \frac{V_2 \dot{V}^2_1- 2 V_1 \dot{V}_1 \dot{V}_2}{2V^2_1}  \right] \;\;\; \;\; \;\;
%\ee
%where $\ell_{\Lambda} = 1/\sqrt{\Lambda}$ and $c$ is a dimensionless constant. To understand its physical meaning, recall that the geometry being homogeneous w.r.t. the hypersurface $\Sigma_{\epsilon}$, one has to restrict the integration over $\Sigma_{\epsilon}$ to avoid divergences. This naturally introduces an infrared cut-off for the system fixing its size. Concretely, one has
%\be
%\label{c}
%c = \frac{1}{\ell^3_p} \int^{t_2}_{t_1} \rd t \oint \ell^2_s \rd \Omega = \frac{\ell_0 \ell^2_s}{\ell^3_p}
%\ee
%Thus, this parameter encodes the ratio between the IR scales and the UV scale of the system. 

The lapse $N(r)$ has been completely absorbed in the definition of the radial coordinate $\tau$. We can safely proceed to describing the system's phase space and evolution with respect to this coordinate.
This is equivalent to gauge-fixing the lapse to $N=\sqrt{\alpha/2\beta}$.
We must nevertheless retain the equation of motion corresponding to lapse variations $\delta N$, which implies that the Hamiltonian vanishes, as customary for relativistic systems. Solving the field equations gives the metric
\be
\rd s^{2}=-\eps \f{\alpha}{2\beta}\rd \tau^{2}+\eps\f{2\beta}{\alpha}\rd t^{2}+\ell_{s}^{2}\alpha\rd\Omega^{2}\,,
\ee
with $\alpha=k^{2}(\tau-\tau_{0})^{2}$ and
%\be
%\f{2\beta}{\alpha}=\f{-\eps}{k^{2}\ell_{s}^{2}}\left[
%1-\f{(\tau_{1}-\tau_{0})}{(\tau-\tau_{0})}-\f{k^{2}\ell_{s}^{2}}{3\ell_{\Lambda}^{2}}(\tau-\tau_{0})^{2}
%\right]
%\,,
%\ee
%
\be
-2\eps\beta
=
\f{1}{\ell_{s}^{2}}(\tau-\tau_{0})(\tau-\tau_{1})
-
\f{k^{2}}{3\ell_{\Lambda}^{2}}(\tau-\tau_{0})^{4}
\,,
\ee
where $\tau_{0}$, $\tau_{1}$ and $k$ are constants of integration.
Rescaling the coordinates as $r=k\ell_{s}(\tau-\tau_{0})$ and $\tilde{t}=t/k\ell_{s}$, we recover the  Schwarzschild-(A)dS solutions,
\be
\rd s^{2}=-f(r)\rd\tilde{t}^{2}+f(r)^{-1}\rd r^{2}+r^{2}\rd\Omega^{2}
\,,
\ee 
with the metric component,
\be
f(r)=1-\f{\ell_{M}}{r}-\f{r^{2}}{3\ell_{\Lambda}^{2}}
\quad\textrm{with}\,\,
\ell_{M}=k\ell_{s}(\tau_{1}-\tau_{0})
\,.\quad
\ee
The constant of integration $\tau_{0},\tau_{1}, k$ and the IR regularization scale $\ell_{s}$ are combined together into the single physical parameter $\ell_{M}$, which gives the Schwarzschild mass of the black hole.

\smallskip

In order to study the symmetries of black hole mechanics, it is convenient to switch to its phase space description. 
We compute the canonical momenta:
\be
\po = \frac{\epsilon c \ell_P}{\vo^{2}} (\vt \dvo - \vo \dvt)
\;, \qquad
\pt = - \epsilon c \ell_P \frac{\dvo}{\vo}
\;,
\ee
forming the canonical pairs $\{\vo,\po\} = \{\vt,\pt \} =1$. The Hamiltonian reads
%\be
%H
%=
%- \frac{1}{\epsilon c \ell_P} \left[ V_1 P_1 P_2 + \frac{1}{2} V_2 P^2_2\right] - \frac{c\ell_P}{\ell^2_{\Lambda}} V_1 -  \frac{c \ell_P}{\ell^2_s}
%\ee
\be
\H=\bHl-\f{c\ell_{P}}{\ell_{s}^{2}}
\quad\textrm{with}\;\;
\bHl=\bH^{(0)}+ \frac{c\ell_P}{\ell^2_{\Lambda}} \vo\;,
\ee
\be
\textrm{and}\quad
\bH^{(0)}=- \frac{1}{\epsilon c \ell_P} \left[ \vo\po\pt + \frac{1}{2} \vt\pt^{2}\right]
\;.
\ee
Remember that we need to impose that the Hamiltonian vanishes $\H=0$. This Hamiltonian constraint consists in a kinetic term $\bH^{(0)}$, a potential term whose coupling is the cosmological constant and a constant shift. This constant shift depends on the IR/UV ratio $c$. It is crucial, since it changes the on-shell value of $\bHl$.
%
%Then, imposing both the Hamilton equations and $\H=0$, one recovers the standard Schwarzschild-(A)dS solutions.
%Notice that the last term of the hamiltonian is a constant shift. As we shall see, it will be convenient to introduce the shifted hamiltonian $\tilde{H} = H + c\ell_P/\ell^2_s$.
%
Now that the dynamics of black holes has been formulated as a mechanical  system, let us show that it admits a symmetry group isomorphic to the Schr\"odinger group.
%
%We are now ready to present the Schrodinger symmetry of this system.

%\rev{{\bf [E] Should give classical solutions! especially identification of MASS. A paper says $\vo\po$, another says $\vo^{\f32}\po\pt$\dots}}

%%%
\section{Schr\"odinger charges for Black Holes}
%%%

As static spherically symmetric metrics in general relativity have been recast as a mechanical system with two degrees of freedom, we  expect a symmetry under the $d=2$  Schr\"odinger group, if it were a free system. The potential actually vanishes when the cosmological constant is set to 0, or equivalently when the cosmological scale is sent to infinity, $\ell_{\Lambda} \rightarrow + \infty$. In that case, we naturally identify Schr\"odinger charges. Below, we further show that, surprisingly, the cosmological potential does not spoil this symmetry, and so the Schr\"odinger group still drives the black hole dynamics whatever the value of $\Lambda$.

\smallskip

Let us start with the case $\ell_{\Lambda} \rightarrow + \infty$, corresponding to a vanishing cosmological constant $\Lambda=0$ and  asymptotically flat Schwarzschild black holes.
Symmetries are generated by conserved charges $\O$, here satisfying
\be
\rd_{\tau}\O=\pp_{\tau}\O+\{\O,\bH^{(0)}\}=0\,.
\ee
Time-independent charges, i.e. with $\pp_{\tau}\O=\{\O,\bH^{(0)}\}=0$, correspond to conformal Killing vectors in the field configuration space $(\vo,\vt)$, while explicitly time-dependent charges, i.e. $\pp_{\tau}\O\ne0$, correspond to conformal Killing vectors in an extended field configuration space given by the Eisenhart-Duval lift \cite{Cariglia:2016oft}. This general approach was pushed forward in \cite{BenAchour:2022fif} to investigate symmetries of gravitational mini-superspaces.
%
%\noindent
Here, we identify translation and boost charges:
%
%\begin{align}
%P_+&=2\sqrt{\vo}\po+\frac{\vt}{\sqrt{\vo}}\pt \\ 
%P_-&=2\sqrt{\vo}\pt\\ %+\frac{1}{L_\Lambda^2}\left[2 c\tau	\sqrt{V_1} +\frac{\tau^2\epsilon \sqrt{V_1}}{2}P_2  \right]\\
%B_+&= \frac{2\epsilon c \vt}{\sqrt{\vo}}+ \tau\left(2\sqrt{\vo} \po +\frac{\vt}{\sqrt{\vo}}\pt\right) \\%+\frac{\tau}{L_\Lambda^2}\left[ c \tau\sqrt{V_1} +\frac{\tau^2 \epsilon \sqrt{V_1}}{6}P_2\right]\\
%B_-&=4\epsilon c\sqrt{\vo}+2\tau \sqrt{\vo} \pt
%\end{align}
%\be
%\begin{split}
%P_+&=\sqrt{\vo}\po+\frac{\vt\pt}{2\sqrt{\vo}} \,,\\
%P_-&=\sqrt{\vo}\pt\,,
%\end{split}
%\quad
%\begin{split}
%B_+&= \eps\frac{\vt}{\sqrt{\vo}}+ \f\tau{c\ell_{P}} P_{+}\,,\\
%B_-&=\eps 2 \sqrt{\vo}+\f\tau{c\ell_{P}} P_{-}\,.
%\end{split}
%\ee
\be
\begin{split}
P_+&=\sqrt{\vo}\po+\frac{\vt\pt}{2\sqrt{\vo}} \,,\\
P_-&=\sqrt{\vo}\pt\,,
\end{split}
\quad
\begin{split}
c\ell_{P}B_+&= \eps c\ell_{P}\frac{\vt}{\sqrt{\vo}}+\tau P_{+}\,,\\
c\ell_{P}B_-&=\eps c\ell_{P}2 \sqrt{\vo}+\tau P_{-}\,.
\end{split}\qquad
\ee
They form a closed Lie algebra with the charge $J = 2\vo\po$:
%\be
%\{P_{-},P_{+}\}=\{B_{-},B_{+}\}=0
%\,,
%\ee
%\be
%\{B_{\pm},P_{\pm}\}=0
%\,,\quad
%\{B_{\pm},P_{\mp}\}=4\eps c \ell_{P}
%\,,
%\ee
%\be
% \left\{ J , B_\pm   \right\}= \pm \frac{1}{2} B_{\pm}
% \,,\quad
% \left\{ J , P_\pm   \right\}= \pm \frac{1}{2} P_{\pm}
% \,.
%\ee
\be
 \begin{split}
&\{P_{-},P_{+}\}=0 \,,\\
&\{B_{\pm},P_{\pm}\}=0\,,\\
 &\left\{ J , B_\pm   \right\}= \pm  B_{\pm}\,,
 \end{split}
 \quad
 \begin{split}
&\{B_{-},B_{+}\}=0 \,,\\
&\{B_{\pm},P_{\mp}\}=\eps\,,\\
&\left\{ J , P_\pm   \right\}= \pm  P_{\pm} \,,
\end{split}
\ee
where $J$ generates $\so(1,1)$ boosts.
We recognize the algebra of Galilean symmetries in two dimensions.
%a centrally extended Heinsenberg algebra.
%
Further introducing the dilatation generator  $D = (\vo\po+\vt\pt)$, we complete this set of conserved charges with the following observables, 
\begin{align}
&&Q_{+} = c\ell_{P} \bH^{(0)}
\,,\quad
Q_{0}  =  D- \tau \bH^{(0)}
\,,\\
&&c\ell_{P}Q_{-}  =  - 2 \epsilon c \ell_P  \vt- 2 \tau D +\tau^2   \bH^{(0)}
 \,,\nn
\end{align}
which form a $\sl(2,\mathbb{R})$ Lie algebra,
\begin{align}
\label{sl}
  \{ Q_0, Q_{\pm}\} = \pm Q_{\pm} \;,  \qquad   \{ Q_{+}, Q_{-}\} = -2 Q_0 
%  \{ Q^{\epsilon}_0, Q^{\epsilon}_{\pm}\} = \mp Q^{\epsilon}_{\pm} \;,  \qquad   \{ Q^{\epsilon}_{+}, Q^{\epsilon}_{-}\} = 2 Q^{\epsilon}_0 
\end{align}
The two sectors are coupled by non-vanishing Poisson brackets:
\be
\label{eqn:coupling}
\begin{split}
&\left\{ Q_{0}, P_\pm   \right\}=\frac{1}{2}P_\pm\,,
\\
& \left\{  Q_{-}, P_\pm  \right\}=-B_\pm \,,
\end{split}
\quad
\begin{split}
&\left\{  Q_{0}, B_\pm  \right\}=-\frac{1}{2}B_\pm \,,
\\
& \left\{ Q_{+}, B_\pm   \right\}=P_\pm\,,
\end{split}
\ee
leading to the 2d centrally extended Schr\"{o}dinger algebra
%$\sh(2) = (sl(2,\mathbb{R}) \times so(2)) \ltimes ( \mathbb{R}^2 \times \mathbb{R}^2) $. 
$\schr(2) = (\sl(2,\mathbb{R}) \oplus \so(1,1)) \oplus_{s} (\mathbb{R}^2 \oplus \mathbb{R}^2)$.
The fact that we have the symmetry subalgebra $\so(1,1)$ instead of $\so(2)$  comes from working with a relativistic system, whose kinetic terms have a non-positive signature. This will become clearer below when diagonalizing explcitily the kinetic terms.
Its quadratic and cubic Casimir both vanish, as expected in classical mechanics:
\be
\C_{2}
=
P_{+}B_{-}-P_{-}B_{+}-\eps J
=0
\,,
\ee
\beq
\C_{3}
&=&
Q_{0}^{2}-Q_{+}Q_{-}-\f14 J^{2}
-\eps B_{+}B_{-}Q_{+}-\eps P_{+}P_{-}Q_{-}
\nn\\
&&-\eps (B_{-}P_{+}+B_{+}P_{-})Q_{0}
+\tfrac\eps2(B_{-}P_{+}-B_{+}P_{-})J
=
0
\,.\nn
\eeq
The latter is the Schr\"odinger-invariant expression of the balance equation for the $\sl_{2}$ Casimir,
\be
Q_{0}^{2}-Q_{+}Q_{-}=\f14 J^{2}\,.
\ee
It is interesting to notice that the evolving position observables $B_{\pm}$ allow to define a canonical transformation to phase space coordinates that diagonalize the kinetic Hamiltonian. Indeed, we read position coordinates from $B_{pm}(\tau=0)$:
\be
X_{+}=\beta/\sqrt{\alpha}
\,,\quad
X_{-}=2\sqrt{\alpha}
\,,\quad
\{X_{\mp},P_{\pm}\}=1
\,.
\ee
Now the Hamiltonian takes a very simple form,
\beq
\bHl
&=&
-\f{\eps}{c\ell_{P}}P_{-}P_{+}
+
\f{c\ell_{P}}{4\ell_{\Lambda}^{2}}X_{-}^{2}
\\
&=&
\f{\eps}{2c\ell_{P}}(P_{2}^{2}-P_{1}^{2})
+\f{c\ell_{P}}{8\ell_{\Lambda}^{2}}(X_{1}+X_{2})^{2}
\,,\nn
\eeq
where we have introduced
\be
P_{\pm}=\f{P_{1}\pm P_{2}}{\sqrt{2}}
\,,\qquad
X_{\pm}=\f{X_{1}\mp X_{2}}{\sqrt{2}}
\,.
\ee
This clarifies the mapping of black hole mechanics onto the $d=2$ particle, with the awkward sign switch in the kinetic term, here $(P_{2}^{2}-P_{1}^{2})$ instead of $(P_{2}^{2}+P_{1}^{2})$.
This sign is a central feature of general relativity: we are working with a 1+1-d relativistic particle.
The non-positive signature signals the gravitational instability (due to conformal factor) that leads to gravitational collapse, black holes and cosmological expansion.
% (see e.g. \cite{Padmanabhan:2016lul}).
%
The black hole phase space IR/UV ratio $c$ plays the role of the 1+1-d particle mass. Keep in mind that the black hole mass is a variable in black hole mechanics. It is a property of the chosen classical solution. More precisely, it is actually a conserved quantity, which we express in terms of the Schr\"odinger charges below in \eqref{eqn:BHmass}.
The cosmological constant creates a quadratic trapping potential for the center-of-mass of the system. As this is a quadratic potential, it seems that one could absorb it in a redefinition of the momenta. It is indeed what happens, as we show below. This is our main result.
%
%Then, the IR/UV ratio $c$ plays the role of the mass, while the cosmological constant creates a quadratic trapping potential for the center-of-mass of the system. As this is a quadratic potential, it seems that one could absorb it in a redefinition of the momenta. It is indeed what happens, as we show below. This is our main result.

\smallskip

Indeed, turning on the cosmological constant ${\Lambda}\ne 0$, we find, quite remarkably, that the Schr\"{o}dinger algebra is preserved.
The conserved charges are mildly modified.
Explicitly, while $P_{-}$ and $B_{-}$ do not acquire corrections, the other translation and boost charges become
\begin{align}
P^{(\Lambda)}_+&=
P_{+}-\eps\f{c^{2}\ell_{P}^{2}}{\ell_{\Lambda}^{2}}\f{\sqrt{\alpha}}{p_{\beta}}
\,,\\
B^{(\Lambda)}_+&=
\frac{\vt^{(\Lambda)}}{\sqrt{\vo}}
+\epsilon \f\tau{c\ell_{P}} P_{+}^{(\Lambda)}
\,,\nn\\
J^{(\Lambda)} & =
J -\eps\f{4c^{2}\ell_{P}^{2}}{3\ell_{\Lambda}^{2}}\f{\alpha}{p_{\beta}}
\,.\nn
%J^{(\Lambda)} & = J - \frac{2}{3} \frac{\epsilon c^{2} \ell^2_p}{\ell^2_{\Lambda}} \frac{\vo}{\pt}
\end{align}
%
%{
%\begin{align}
%P^{(\Lambda)}_+&=
%P_{+}-\f{c^{2}\ell_{P}^{2}}{\ell_{\Lambda}^{2}}\f{\sqrt{\alpha}}{p_{\beta}}
%\,,\\
%B^{(\Lambda)}_+&=
%\frac{\vt}{\sqrt{\vo}}-\f{2c^{2}\ell_{P}^{2}}{3\ell_{\Lambda}^{2}}\f{\sqrt{\alpha}}{p_{\beta}^{2}}
%+\epsilon \f\tau{c\ell_{P}} P_{+}^{(\Lambda)}
%\,,\nn\\
%J^{(\Lambda)} & =
%J -\f{4c^{2}\ell_{P}^{2}}{3\ell_{\Lambda}^{2}}\f{\alpha}{p_{\beta}}
%\,.\nn
%%J^{(\Lambda)} & = J - \frac{2}{3} \frac{\epsilon c^{2} \ell^2_p}{\ell^2_{\Lambda}} \frac{\vo}{\pt}
%\end{align}
%}
The conformal sector is similarly modified,
\beq
&&Q^{(\Lambda)}_{+}  =
c\ell_{P}\bHl
\,,
\quad
Q^{(\Lambda)}_{0} =
D^{(\Lambda)}-\tau\bHl
%-\f{4c^{2}\ell_{P}^{2}}{3\ell_{\Lambda}^{2}}\f{\alpha}{p_{\beta}}
\,,\\
&&
c\ell_{P}Q^{(\Lambda)}_{-} =
 - 2 \epsilon c \ell_P  \vt^{(\Lambda)}- 2 \tau D^{(\Lambda)} +\tau^2   \bHl
%+\f{4c^{2}\ell_{P}^{2}}{3\ell_{\Lambda}^{2}}\left[
%c \ell_P  \f{\alpha}{p_{\beta}^{2}}+2\tau\f{\alpha}{p_{\beta}}
% \right]
\,,\nn
\eeq
with the following $\Lambda$-corrections:
\be
\beta^{(\Lambda)}
=
\beta-\eps\f{2c^{2}\ell_{P}^{2}}{3\ell_{\Lambda}^{2}}\f{\alpha}{p_{\beta}^{2}}
\,,\quad
D^{(\Lambda)}
=
D-\eps\f{4c^{2}\ell_{P}^{2}}{3\ell_{\Lambda}^{2}}\f{\alpha}{p_{\beta}}
\,.\quad
\ee
The new conserved charges satisfy $\pp_{\tau}\O+\{\O,\bHl\}=0$, and the Hamiltonian  simply reads $c\ell_{P}\bHl=-\eps P_{-}P_{+}^{(\Lambda)}$.
We get the same Lie algebra as for the $\Lambda=0$ case.
It follows that the mechanics of Schwarzschild-(A)dS black holes is also invariant under the non-relativistic conformal Schr\"{o}dinger symmetry. 

This result parallels the fact that the Schr\"{o}dinger symmetry for 1d classical mechanics is preserved for two specific potentials: the harmonic potential and the inverse square potential (whose quantization was studied in \cite{deAlfaro:1976vlx}). From that point of view, the Schwarzchild-(A)dS black hole mechanics can be viewed as an extension of the flat Schwarzchild black hole mechanics similar to the extension of the free particle to the harmonic oscillator (with a positive or negative pulsation).

We can compute the value of those observables on classical solutions. In particular, we get:
\be
J^{(\Lambda)}=\f{c\ell_{P}}{\ell_{s}^{2}}(\tau_{1}-\tau_{0})
\,,\quad
P_{-}=-\eps 2c\ell_{P}k
\,,
\nn
\ee
which allows to identify the black hole mass as a conserved charge:
\be
\label{eqn:BHmass}
\ell_{M}=-\eps\f{2\ell_{s}^{3}}{c^{2}\ell_{P}^{2}}\,J^{(\Lambda)}P_{-} ^{(\Lambda)}
\,.
\ee
It is definitely intriguing that the black hole mass is equal to the boost generator $J^{(\Lambda)}$ (up to the velocity factor $P_{-}$). This evocates recent discussions on black hole thermodynamics where Hawking's thermal radiation is derived from the identification of the (modular) Hamiltonian as a boost generator within a $\sl(2,\R)$ algebra of asymptotic symmetry generators (see e.g. \cite{Jafferis:2020ora}). Although the symmetry structure is very similar, the link between the present work and this framework is not obvious at all.

An important remark is that the cosmological constant  $\ell_{\Lambda}$ never appears in the on-shell values of the Schr\"odinger charges.
For instance, the cosmological constant does not change the Schr\"odinger Casimirs $\C_{2}=\C_{3}=0$. In fact, $\Lambda$ shifts the definition of the conserved charges but does not affect at all the Schr\"odinger symmetry. Let us insist that these are not space-time isometries or diffeomorphisms, but non-trivial symmetry of general relativity under transformations acting on the space of metrics.

Here, we have found  that the cosmological constant does not affect the symmetry of general relativity, at least in the spherically symmetric sector. From the point of view of symmetries, $\Lambda$ will appear back when breaking the Schr\"odinger symmetry, for instance by introducing an ``observer''. This can be simply achieved by going beyond the gravitational sector and looking at the dynamics of matter fields coupled to the geometry, in case the cosmological constant will surely modify the dynamics and symmetries of the matter field evolution.
%
%For instance, the cosmological constant does not change the Schr\"odinger Casimirs $\C_{2}=\C_{3}=0$. In fact, $\Lambda$ shifts the definition of the conserved charges but does not affect at all the Schr\"odinger symmetry. This means that the cosmological constant does not affect the symmetry of general relativity, at least in the spherically symmetric sector. From the point of view of symmetries, $\Lambda$ will appear back when breaking the Schr\"odinger symmetry by introducing an ``observer''. This can be simply achieved by going beyond the gravitational sector and looking at the dynamics of matter fields coupled to the geometry, in case the cosmological constant will surely modify the dynamics and symmetries of the matter field evolution.

%%%%%%%%%%
\section{Discussion \& Prospects}
%%%%%%%%%%

We have shown that the dynamics of stationary spherically-symmetric metrics in general relativity can be formulated as a two-dimensional mechanical system with a non-trivial potential whose coupling constant is the cosmological constant. We call this model black hole mechanics. Keep in mind that the evolution parameter here is the radial coordinate, which is space-like outside the black hole and time-like in the interior region.
This allowed us to show that the black hole mechanics is invariant under the $d=2$ Schr\"odinger group. This invariance holds both for the interior and the exterior regions of the black hole. Moreover, it holds whatever the value of the cosmological constant $\Lambda$. The symmetry transformations act on the phase space of geometries and are not mere space-time transformations. They change the black hole mass $\ell_{M}$ and the singularity position $\tau_{0}$, as well as the IR regularization scale $\ell_{s}$, while leaving the equations of motion invariant. 

Since the  Schr\"odinger group is the key (maximal) symmetry of classical mechanics which is preserved under quantization, it is natural to expect quantum black holes to retain this symmetry. Breaking this symmetry when quantizing black holes would definitely signal a strong deviation with respect to the standard quantization logic and would reveal some important hidden physical ingredients in the description of black holes in general relativity. 

Digging deeper in this direction, here we have taken the perspective of considering quantum mechanics as a field extension of classical mechanics. Putting aside conceptual issues (e.g. the measurement problem and collapse of the wave-function), quantum mechanics is mathematically formulated as a description of the dynamics of the wave-function: classical positions and momenta, evolving in time, are replaced by a wave-function, considered as a space-time field, interpreted as a dressed classical object with classical positions and momenta, plus extra degrees of freedom representing the shape fluctuations of the wave packet.

From this view point of quantization as field extension and turning to black holes, there are actually two natural field extensions of black hole mechanics:
\begin{itemize}
\item
On the one hand, it is natural to quantize black hole mechanics and lift a classical black hole metric to a wave-function with a fuzzy mass and a fuzzy singularity. Let us underline that this does not mean relaxing the hypothesis of stationarity or spherical symmetry: we describe quantum superposition of spherically symmetric metrics. This goes in the same direction as the line of research on effective black hole metrics taking into account quantum gravity corrections and attempting to solve the singularity problem without introducing anisotropy or leaving spherical symmetry, e.g. \cite{Kuchar:1994zk, Bojowald:1999ex, Sartini:2021ktb}.
Our analysis means that preserving the Schr\"odinger symmetry should be crucial to this approach (see e.g. \cite{BenAchour:2018jwq} using the conformal symmetry to constrain regularized black hole metrics in effective quantum gravity models).

\item
On the other hand, the natural field theory of black hole mechanics is general relativity, which reestablishes inhomogeneities and anisotropies on top of the spherically symmetric background and describes their dynamics. From this perspective, general relativity is to be interpreted more as the non-perturbative field theory of black hole excitations, instead of its usual interpretation as the field theory encoding the non-linear properties of gravitational waves.
More precisely, general relativity would lead to a non-perturbative hydrodynamic description of the black holes microstates
%(possibly augmented by further microscopic dofs),
with the black hole sector identified by the Schr\"odinger symmetry we have found here.
Then it would be natural to understand if general relativity is invariant under an extension of the Schr\"odinger symmetry group. Let us underline that we expect that these symmetries would not be space-time diffeomorphisms, but non-trivial transformations on the phase space of geometries.
Interestingly, it has been recently shown that the static perturbations of the Schwarzschild and Kerr black holes relevant to compute the Love numbers are also governed by a Schr\"{o}dinger symmetry \cite{BenAchour:2022uqo}. It would be interesting to further understand how this symmetry for perturbations can be related to the background symmetry discussed here.
From a more general perspective, it would be enlightening to compare the Schr\"odinger charges derived here to the existing extended BMS charges and $w_{1+\infty}$ charge algebra for asymptotically flat space-time
as derived in e.g.
\cite{Strominger:2021lvk,Freidel:2021ytz,Fuentealba:2022gdx,Barnich:2022bni}.

\end{itemize}

For both field theory extensions of black hole mechanics, we expect the Schr\"odinger Casimirs, $\C_{2}$ and $\C_{3}$,  not to vanish anymore, and to reflect the extra structures and degrees of freedom dressing the black hole evolution. If both quantum black holes and asymptotically flat general relativity turn out both to preserve the Schr\"odinger symmetry, it should definitely be a {\it key symmetry of quantum gravity}.

\medskip

As a direct application of the present work, we would like to point out that  the Schr\"{o}dinger symmetry can be used to select quantum corrections to black hole dynamics.
Indeed, now the canonical analysis of the classical black hole dynamics has been clarified, it is straightforward to quantize the system. We define quantum states as wave-functions of the metric components. Instead of using the original variables $\alpha,\beta$, our analysis suggests that using the variable $X_{1},X_{2}$, and thus considering wave-functions $\Psi(X_{1},X_{2})$, is more convenient. Then this wave-function is driven by a field action:
\be
S[\Psi,\bPsi]
=
\int \rd\tau\,\Big{[}
i\hbar\bPsi\pp_{\tau}\Psi-\bPsi\widehat{\bHl}\Psi
\Big{]}\,,
\ee
where the Hamiltonian operator $\widehat{\bHl}$ consists in a kinetic operator, given by the  $1+1$-dimensional Laplacian,  plus a harmonic potential term whose coupling constant is given by the cosmological constant. The analysis of the symmetry of quantum mechanics, reviewed in section \ref{sec:QM}, shows that the Schr\"odinger symmetries are preserved by this standard quantization scheme. 
Just as atom-atom microscopic interactions in quantum mechanics can be taken into account by introducing a potential $\V[\Psi, \bar{\Psi}]$, which encodes the self-interaction of the wave-function fluctuations and excitations, e.g. \cite{Kolo, Ghosh:2001an}, we similarly expect that quantum gravity will lead to a self-interaction between the quanta of geometry forming the black hole, thus leading to an effectively modified field action:
\be
S_{QG}
=
\int \rd\tau\,\Big{[}
i\hbar\bPsi\pp_{\tau}\Psi-\bPsi\widehat{\bHl}\Psi
+\V[\Psi, \bar{\Psi}]
\Big{]}\,.
\ee
The key point is that the Schr\"{o}dinger charge algebra also holds for precise non-linear extensions of the Schr\"{o}dinger dynamics.
%
%Another point which will become relevant in the following is that this Schr\"{o}dinger charge algebra also holds for suitable non-linear extensions of the Schr\"{o}dinger dynamics \cite{Kolo, Ghosh:2001an}. Indeed, one can deform the hamiltonian by a self-interacting potential $V(\Psi, \bar{\Psi}, \vec{x})$, introducing thus a new scale $\gamma$ encoding the atom-atom microscopic interaction such that
%\be
%Q_{+} = \int \rd^d x \left( \bar{\Psi} \partial_i \partial^i \Psi  + \gamma V(\Psi, \bar{\Psi}, \vec{x}) \right)
%\ee
%
Remarkably, depending on the spatial dimension $d$, one can show that the Schr\"{o}dinger charge algebra is preserved for suitable self-interaction. Since such a potential is homogeneous, it does not affect the symmetry under phase transformations, translations and boosts. And one easily checks  that the conformal symmetry \eqref{eq:conformal} is preserved for $\V\propto |\Psi|^{2n}$ when $d(n-1)=2$.
For $d=1$, this selects a self-interaction potential $\V\propto |\Psi|^6$ leading to the Tonks-Girardeau equation, which defines a quintic non-linear extension of the Schr\"odinger equation.
Here, the black hole mini-superspace model corresponds to $d=2$ dimensions, and preserving the full Schr\"odinger symmetry selects a quadratic potential $\V \propto |\Psi|^4$:
\be
S_{QG}^{\kappa}
=
\int \rd\tau\,\Big{[}
i\hbar\bPsi\pp_{\tau}\Psi-\bPsi\widehat{\bHl}\Psi
-\kappa|\Psi|^4
\Big{]}\,.
\ee
%
%This is the case for the Gross-Pitaevskii equation in $d=2$ dimensions with a quadratic potential $\V \propto |\Psi|^4$, or the Tonks-Girardeau equation for $d=1$ with the self-interaction potential $\V\propto |\Psi|^6$ leading to the quintic non-linear Schr\"odinger equation.
This shows that there exists a non-trivial UV corrected quantum dynamics protected by the Schr\"odinger symmetry.
The evolution of the black hole wave-function is then driven by a Gross-Pitaevskii equation,
\be
i\hbar\pp_{\tau}\Psi=\widehat{\bHl}\Psi
+2\kappa|\Psi|^2\Psi\,.
\ee
The new coupling $\kappa$ controls the attraction or repulsion between wave-packets depending on its sign. It will thus determine when and where one can have stable quantum superpositions of black hole states, and  enter in a crucial fashion in the identification of the transitions between the semi-classical regime and the deep quantum regime.

Playing with this new parameter $\kappa$ should lead to a new phenomenology for quantum black holes. It is very different from modified gravity theories, which usually propose modifications of the Hamiltonian operator but do not consider non-linear self-interaction terms.
We consider such symmetry-protected non-linear extension of the Wheeler-de Witt equation as a universal template for the effective dynamics of quantum black holes. It would be enlightening to understand which quantum gravity models or scenarii generate such quantum black hole dynamics. 

\medskip

We would like to conclude with the  remark that, using  the phase-amplitude factorization of the wave-function $\Psi = \sqrt{\rho} e^{i\theta}$, the Schr\"odinger equation and its non-linear extension can be understood as Navier-Stokes' equation for compressible fluid dynamics leading to the hydrodynamics reformulation of quantum mechanics, e.g. see the original seminal work by Madelung \cite{Madelung} and the more recent analysis \cite{Horvathy:2009kz}. Since black hole mechanics is invariant under the $d$=2 Schr\"odinger group, this means that we get an intriguing mapping between black hole quantum mechanics and two-dimensional hydrodynamics. It is tempting to speculate that this could be related to a fluid dynamics for gravitational quanta on the black hole horizon considered as a 2d membrane (as in the corner dynamics for general relativity \cite{Donnelly:2020xgu}). A more down-to-earth expectation is this mapping surely provides a promising avenue to reformulate quantum black hole as a many-body Schr\"{o}dinger system.
In fact, it opens the door to the possibility of a new class of analogue condensed matter models for black holes, cosmology and quantum gravity phenomenology, based on an exact mapping of symmetries, conserved charges and dynamics, instead of focusing on shaping and manufacturing equivalents of space-time metrics.
%
%Notice that in contrast with previous works \cite{Louko:1996md, Berezin:1998xf, Neronov:1998qc, Vaz:1998gm, Dreyer:2002vy, Krause:2002we}, our derivation is based solely on the symmetries of the homogeneous sector of spherically symmetric vacuum General Relativity. As such, our approach is free from any exotic inputs. In particular, the underlying symmetry discussed here might well be deformed or even broken in the deep quantum regime. However, being a property of the classical theory, it should be regarded as an IR symmetry which has to be recovered in the semi-classical limit. From that perspective, this IR symmetry implies the discreteness of the Schwarzschild mass spectrum at least for sufficiently large semi-classical black hole. 

\bigskip

%\pagebreak[5]

\noindent\textit{\textbf{Acknowledgements.}}
The work of J. Ben Achour is supported by the Alexander von Humboldt foundation and the Sir John Templeton foundation.

\bibliographystyle{bib-style} 

\bibliography{SchAdS}

\end{document}